\documentclass[final]{svjour3}
\usepackage[dvipdfmx]{graphicx}
\usepackage{rotating}
\usepackage{amssymb}
\usepackage{mathptmx}
\usepackage[numbers]{natbib}
\usepackage{wrapfig}
\usepackage[dvipdfmx]{color}

\makeatletter
\journalname{Journal of Low Temperature Physics}
\bibpunct{}{}{,}{s}{}{,}

\begin{document}

\newcommand{\hdblarrow}{H\makebox[0.9ex][l]{$\downdownarrows$}-}
\title{Mechanical cryocooler noise observed in the ground testing of the
\textit{Resolve} X-ray microcalorimeter onboard XRISM}
\titlerunning{Mechanical cryocooler noise in \textit{Resolve} X-ray microcalorimeter}

\author{R. Imamura, H. Awaki, M. Tsujimoto, S. Yamada, F. S. Porter, C. A. Kilbourne, R. L. Kelley, Y. Takei, on behalf of the XRISM
\textit{Resolve} team}
\authorrunning{Imamura et al.}

\institute{Department of Physics, Ehime University, Matsuyama, Ehime, 790-8577, Japan\\
\email{imamura@astro.phys.sci.ehime-u.ac.jp}}

\date{Received: date / Accepted: date}

\maketitle

\begin{abstract}
 Low-temperature detectors often use mechanical coolers as part of the cooling chain in
 order to reach sub-Kelvin operating temperatures. The microphonics noise caused by the
 mechanical coolers is a general and inherent issue for these detectors. We have
 observed this effect in the ground test data obtained with the \textit{Resolve}
 instrument to be flown on the XRISM satellite. \textit{Resolve} is a cryogenic X-ray
 microcalorimeter spectrometer with a required energy resolution of 7 eV at 6 keV. Five
 mechanical coolers are used to cool from ambient temperature to $\sim$4~K: four
 two-stage Stirling coolers (STC) driven nominally at 15 Hz and a Joule-Thomson cooler
 (JTC) driven nominally at 52 Hz. In 2019, we operated the flight-model instrument for
 two weeks, in which we also obtained accelerometer data inside the cryostat at a
 low-temperature stage (He tank). X-ray detector and accelerometer data were obtained
 continuously while changing the JTC drive frequency, which produced a unique data set
 for investigating how the vibration from the cryocoolers propagates to the detector. In
 the detector noise spectra, we observed harmonics of both STCs and JTC. More
 interestingly, we also observed the low ($<$20 Hz) frequency beat between the 4'th JTC
 and 14'th STC harmonics and the 7'th JTC and the 23--24'th STC harmonics. We present
 here a description and interpretation of these measurements.
\end{abstract}

\section{Introduction}\label{s1}
Low-temperature detectors are cooled by a multi-stage cooling system often containing
mechanical coolers. The propagation of the mechanical vibration to the detector is
potentially a serious source of noise. An example is the X-ray microcalorimeter
instrument (Soft X-ray Spectrometer\cite{mitsuda14,kelley16}; SXS) on board the Astro-H
satellite \cite{takahashi16}. The noise due to vibration from the mechanical coolers
caused unacceptable degradation of the detector performance during ground testing. This
was mitigated by installing a vibration isolation system (VIS) between the mechanical
coolers and the cryostat at a very late stage of the mission\cite{takei18}. After the
launch in February 2016, the SXS achieved a stable and unprecedented performance in
orbit\cite{porter18,leutenegger18} but was suddenly discontinued due to the malfunction
of the spacecraft attitude control system.

The X-Ray Imaging and Spectroscopy Mission (XRISM)\cite{tashiro20} was initiated to
recover Astro-H science and is planned to be launched in 2023. The
X-ray microcalorimeter instrument, named \textit{Resolve}, is built almost to the same
design as the SXS. The VIS was installed from the beginning with some design
changes, including the launch lock mechanism and improved damping performance
\cite{ezoe20, ishisaki22}. To evaluate microphonics, we monitor the mechanical vibration
of the cryostat using accelerometers throughout the ground testing and measure the
transfer function from the accelerometers to the detector.

Here, we present some initial findings obtained in a test conducted in December
2019. 
Later findings are discussed in separate articles\cite{imamura22,hasebe22}. 
We start with
a short description of the \textit{Resolve} instrument relevant to this article and
include references for additional information in \S~\ref{s2}. We then describe the
measurements and the data set in \S~\ref{s3}. We present the result in \S~\ref{s4} and
conclude in \S~\ref{s5}. The measurements are ongoing and a full description will be
published at the conclusion of the ground test program.

\section{Instrument}\label{s2}
The X-ray microcalorimeter for \textit{Resolve}\cite{kilbourne18} is based on
ion-implanted Si thermometers with HgTe x-ray absorbers anchored to a 50~mK heat sink
with a thermal time constant of $\sim$3.5~ms. A total of 36 pixels comprise the detector
array. Individual photons are detected and the energy is reconstructed on board. For
nominal operations, the characteristic values of individual pulses are
downloaded. However, for diagnostic purposes, short periods of continuous time-domain
samples are downloaded either by command (nominally up to 4.096~s) or in the background
(81.92~ms) at a 12.5~kHz sample rate. The time-domain data thus obtained yield power
spectra in the 0.00024~-6.25~kHz range.

%\begin{wrapfigure}[17]{r}{0.60\textwidth}
\begin{figure}[!hbtp]
 \centering
 %\vspace*{-9mm}
 \includegraphics[keepaspectratio,width=0.9\textwidth]{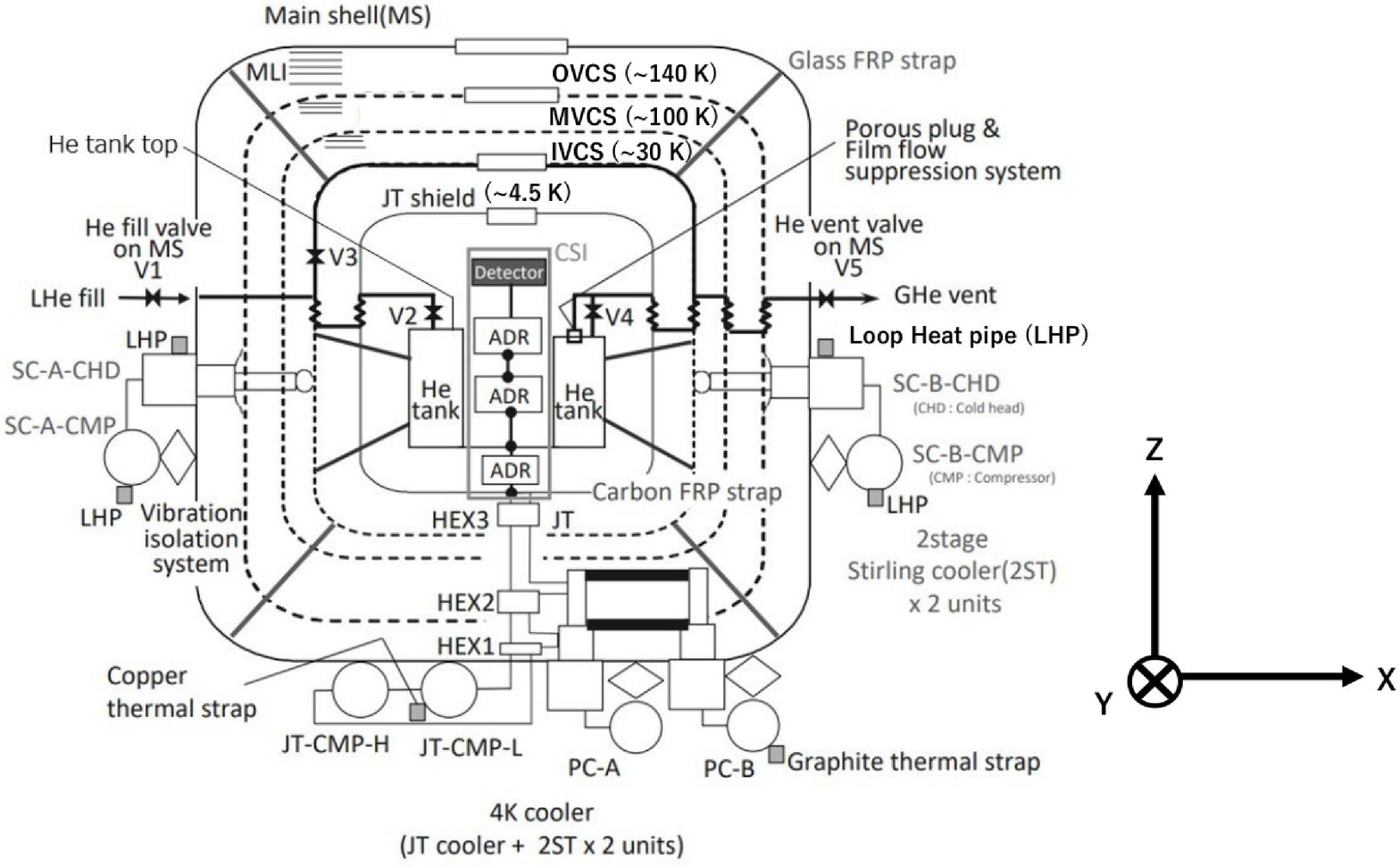}
  \caption{Schematic of the cryostat\cite{ezoe20,yoshida18} and axis definition. Accelerometers were placed at the
 top of the He tank.}
 \label{f03}
%\end{wrapfigure}
\end{figure}

Fig.~\ref{f03} shows the conceptual diagram of the cryostat. The heat sink for the detector is
PID-controlled and fed back to a two-stage adiabatic demagnetization refrigerator
(ADR)\cite{shirron18}. The detector assembly \cite{chiao18} and the ADR stages are
installed in the cryostat\cite{ezoe20,yoshida18,fujimoto18}. The He tank stores
30~liters of superfluid He surrounded by several layers of shields: Joule-Thomson shield
(JTS), inner, middle, and outer vapor cooled shields (IVCS, MVCS, and OVCS,
respectively), and the Dewar main shell (DMS). The shields are mechanically suspended by
straps \cite{fujimoto18}. The He is filled and vented through plumbing with access ports
and valves. When the liquid He is depleted, the system continues to operate in a
cryogen-free mode utilizing a third ADR\cite{sneiderman18}.

Two sets of two-stage Stirling coolers (STC) and one Joule-Thomson cooler (JTC) are
used\cite{sato12}; one STC set (SC-A and B) is for cooling the OVCS and IVCS and the JTC
for cooing the JTS. The other STC set (PC-A and B) is used for pre-cooling the JTC. The
VIS is installed between the cryostat and each one of the compressors of the four STCs
attached to the DMS. The VIS was not installed for JTC as it is known to produce
line-dominated vibration noise spectra, which can be adjusted later by the change of the
drive frequency. The four STCs are nominally driven at $f_{\mathrm{STC}}=15.0$~Hz and
the JTC at $f_{\mathrm{JTC}}=52.0$~Hz by the cooler driver electronics. To avoid
possible noise interference in the instrument, the drivers can change the frequency from
13.93--16.26~Hz with 256 steps at a resolution $\Delta f_{\mathrm{STC}}=0.01$~Hz for the
STCs and from 50.4--53.8~Hz with 32 steps at $\Delta f_{\mathrm{JTC}}=0.1$~Hz for the
JTC.

\section{Measurements}\label{s3}
A cryogenic test was conducted in 2019 December using the flight-model cryocoolers and
the partially assembled cryostat including the flight model detector assembly and ADRs
in order to verify that the hardware was ready for the final integration. The test was
conducted in the Niihama Works of Sumitomo Heavy Industries for two weeks. A three-axis
set of cryogenic accelerometers was temporarily installed on the He tank for this
test. We measured $x$, $y$, and $z$ axis acceleration at the top of the He tank using a
set of model 876 accelerometers from Columbia Research Industries, Inc. The
accelerometers were sampled at 10~kHz with 32~s records, continuously, to cover the
frequency range of 31.25~mHz--5~kHz with a dynamic range of $\sim$100~dB and a noise
floor of 2--3$\times$10$^{-5}$~G/$\sqrt{\mathrm{Hz}}$.

We operated the instrument in the nominal on-orbit configuration; i.e., He filled in the
He tank, cryocoolers operated at nominal power, and the VIS launch locks released. The
detector performance was stable. We examined the detector response using a fixed
$f_{\mathrm{STC}}=$15.0 Hz while varying $f_{\mathrm{JTC}}$ on December 11. We first
moved from $f_{\mathrm{JTC}}=52.0$~Hz upward to 53.8~Hz, and then restarted from 52.0~Hz
and moved downward to 50.4~Hz. At each step, we dwelled for 5 minutes and measured the
detector noise spectra and the accelerometer spectra.

\section{Results}\label{s4}
\begin{figure}[!hbtp]
 \centering
 \includegraphics[width=\columnwidth,clip]{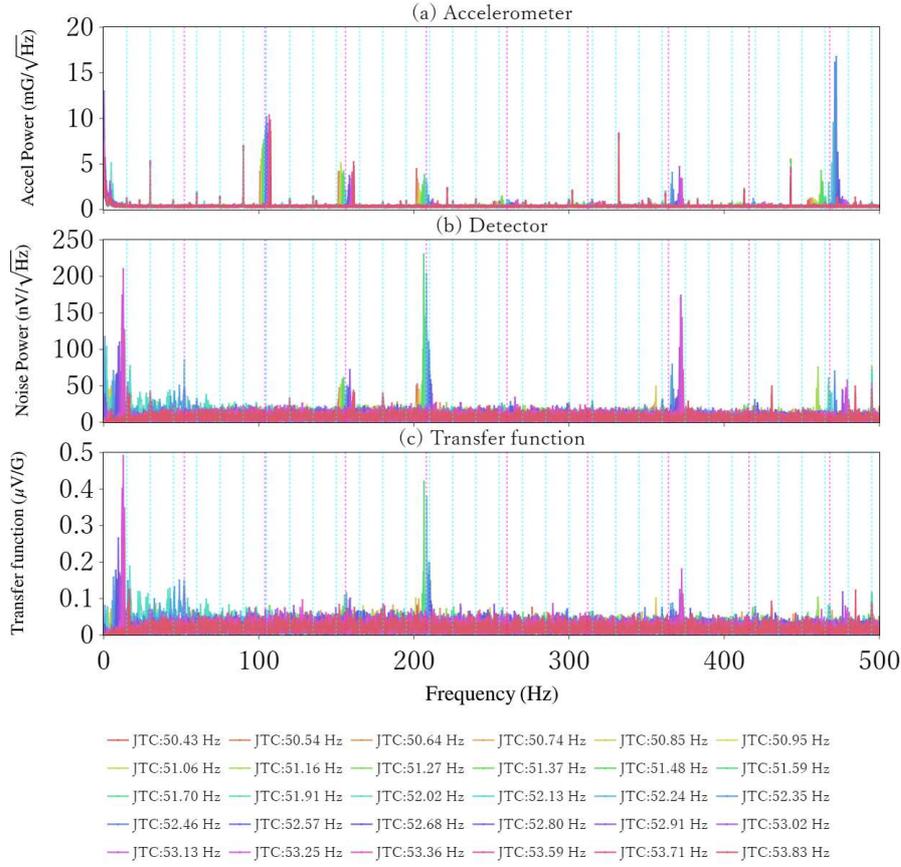}
 \caption{(a) Accelerometer spectra in the $z$-axis of the He tank. (b) Detector noise
 spectra for pixel 0. (c) Transfer function derived from (b)/(a) by matching the
 frequency resolution. Different colors indicate different $f_{\mathrm{JTC}}$ settings. The fundamental and harmonics of $f_{\mathrm{STC}}=15.0$~Hz and
 $f_{\mathrm{JTC}}=52.0$~Hz are shown with dotted cyan and magenta lines, respectively.}
 \label{f01}
\end{figure}

Fig.~\ref{f01} shows (a) the accelerometer spectra in the $z$-axis of the He tank and
(b) the detector noise spectra for a representative detector pixel (pixel 0) in the
0--500~Hz range. In (a), the fundamental and harmonics of the fixed $f_{\mathrm{STC}}$
and varying $f_{\mathrm{JTC}}$ are observed for multiple orders across the frequency
range. This is not surprising for mechanical coupling from the cryocoolers. However, the
propagation of these into the detector noise in (b) exhibits several interesting
features, which are also observed in other combinations of the accelerometer channels
($x$, $y$, and $z$ of the He tank) and detector channels (pixels 0--35).

First, the frequency of the detector line noise changes as $f_{\mathrm{JTC}}$
changes. This is particularly evident for $n_{\mathrm{JTC}}=$~3, 4, 7, and 9 for the
$n_{\mathrm{JTC}}$'th harmonics. This indicates the JTC origin for these lines. Second, the amplitude of the interference lines is very different between different $n_{\mathrm{JTC}}$'s: $n_{\mathrm{JTC}}=$~2
and 3 are the strongest in the accelerometer spectra, but $n_{\mathrm{JTC}}=$~4 and 7
are the strongest in the detector noise spectra. Third, the detector noise at
$f_{\mathrm{JTC}}$ and below is enhanced only when the JTC-drive frequency is close to 52.0~Hz as shown in
blue in (b). Fourth, line noise is present in the low-$f$ range ($\lesssim$20~Hz) in the
detector noise spectra (b). These effects all originate from the cryocoolers as their frequency and
intensity change as $f_{\mathrm{JTC}}$. However, the low-$f$ lines are absent in the
accelerometer spectra (a).

\begin{figure}[!hbtp]
 \centering
 \includegraphics[width=0.95\columnwidth,clip]{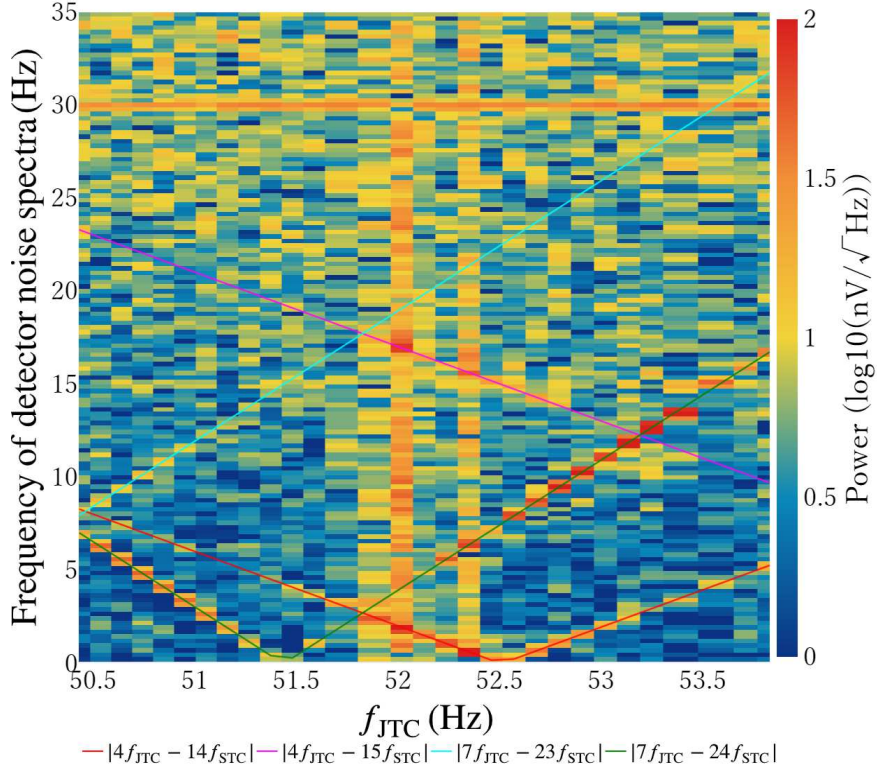}
 \caption{Detector noise spectra (pixel 0) shown in a heat map as a function of
 $f_{\mathrm{JTC}}$. Beat lines calculated by $f_{\mathrm{beat}} = | n_{\mathrm{JTC}}
 f_{\mathrm{JTC}} - n_{\mathrm{STC}} f_{\mathrm{STC}} |$ are shown in different colors.}
 \label{f02}
\end{figure}

Fig.~\ref{f02} is another representation of the third and fourth features discussed
above. The intensity of the detector noise (pixel 0) as a function of $f_{\mathrm{JTC}}$
is shown with a heat map. These data were acquired when $f_{\mathrm{STC}}$ was 15 Hz.
%The noise caused by the STC drive frequency appears as horizontal lines at 30.0~Hz and faintly line at 15~Hz, while those due to the JTC as vertical lines
%most evident at 
The noise caused by the STC drive frequency appears as horizontal lines at 15~Hz (faint) and 30.0~Hz, 
while those due to the JTC as vertical lines are most evident at
$f_{\mathrm{JTC}}=$~52.0--52.4~Hz. 
In addition, we see many diagonal
lines. From the slope, these are associated with $n_{\mathrm{JTC}}=$~4 and
7. Furthermore, from the absolute values of the changing frequencies, we found that they
are the beat frequencies represented by
\begin{math}
 f_{\mathrm{beat}} = | n_{\mathrm{JTC}} f_{\mathrm{JTC}} - n_{\mathrm{STC}} f_{\mathrm{STC}} |,
\end{math}
in which $n_{\mathrm{STC}}=$14 (and 15 for a weak signal) for $n_{\mathrm{JTC}}=4$ and
$n_{\mathrm{STC}}=$24 (and 23 with a weak signal) for $n_{\mathrm{JTC}}=7$. The beat
only appears, in particular, $f_{\mathrm{JTC}}$ ranges. Beat lines for other
$n_{\mathrm{JTC}}$'s were not observed, which is interesting as there is always a
matching $n_{\mathrm{STC}}$ that makes $f_{\mathrm{beat}}<$~20~Hz for all
$n_{\mathrm{JTC}}$'s. 

There should be some physical mechanisms to explain why the beat
frequency lines appear in the detector noise spectra since linear combinations of
$n_{\mathrm{JTC}} f_{\mathrm{JTC}}$ and $n_{\mathrm{STC}} f_{\mathrm{STC}}$ tones cannot
redistribute power into the $f_{\mathrm{beat}}$ tone.
Assuming that $f_1 = n_{\mathrm{JTC}} f_{\mathrm{JTC}}$ and $f_2 = n_{\mathrm{STC}}
f_{\mathrm{STC}}$ tones ($A_1$ and $A_2$, respectively) are linearly mixed with the same
phase and amplitude of unity, the mixture ($A_{12}$) is
\begin{eqnarray}
 A_{12} &=& A_1 + A_2 = \cos{(2\pi f_1 t)} + \cos{(2\pi f_2 t)} = 2 \cos{\left(2\pi\frac{f_1-f_2}{2}t\right)}
  \cos{\left(2\pi\frac{f_1+f_2}{2}t\right)}\nonumber\\ 
  &=& 2 \cos{\left(2\pi f_{\mathrm{mod}}t\right)}
  \cos{\left(2\pi f_{\mathrm{carrier}}t\right)}\nonumber.
\end{eqnarray}
The mixed signal is the amplitude modulation of the single carrier frequency
$f_{\mathrm{carrier}} = (f_1+f_2)/2$ with the modulation frequency $f_{\mathrm{mod}} =
(f_1-f_2)/2 = f_{\mathrm{beat}}/2$. The Fourier transform of $A_{12}$ has power only at
the $f_1$ and $f_2$ tones and not at the $f_{\mathrm{beat}}$ tone. However, when a
non-linear effect takes place on $A_{12}$ to become $A^{\prime}_{12}$, such as a
one-sided clip ($A_{12}^{\prime}=$min($A_{12}$, $A_{12,\mathrm{thres}}$), where
$|A_{12,\mathrm{thres}}|<2$) or a cubic term ($A_{12}^{\prime}= \alpha^{(1)} A_{12} +
\alpha^{(3)} A_{12}^{3}$, where $\alpha^{(1,3)}$ are coefficients), the power is
redistributed into $f_{\mathrm{beat}}$ and its harmonic tones. It is not relevant for
the redistribution whether a non-linear effect takes place in each of $A_1$ or
$A_2$. Only non-linear effect of $A_{12}$ produces the redistribution.

We speculate as follows. When both $f_1$ and $f_2$ are close to resonance frequencies,
their tones are amplified. This only occurs for particular $n_{\mathrm{JTC}}$'s and
$n_{\mathrm{STC}}$'s in order to explain why the beat appears only for
($n_{\mathrm{JTC}}=$4, $n_{\mathrm{STC}}=$14 and weakly 15) and ($n_{\mathrm{JTC}}=$7,
$n_{\mathrm{STC}}=$24 and weakly 23) and in particular $f_{\mathrm{JTC}}$ ranges. In
fact, for these $n_{\mathrm{JTC}}$'s and $n_{\mathrm{STC}}$'s, some resonances are estimated
from mechanical models: 209~Hz for the ADR salt pill and 372 Hz for the detector
assembly. The beat from $n_{\mathrm{JTC}}=4$ (Fig.~\ref{f02}) disappears at
$f_{\mathrm{JTC}}<51.6$~Hz presumably because $4f_{\mathrm{JTC}}$ is too far from
209~Hz. The beat from $n_{\mathrm{JTC}}=7$ appears in two parts: $f_{\mathrm{JTC}}<51.5$
and $>52.0$~Hz, suggesting that there are two beat frequencies at $\sim$355 and
372~Hz. A non-linear effect, though its exact physical mechanism is still unknown, must
take place in the mixed signal of the two amplified monotones, which then redistributes
power into the $f_{\mathrm{beat}}$ tone.  Another possible origin of the
$f_{\mathrm{beat}}$ tone is the sensitivity of the detector on the amplitude of the
modulation $|\cos(2\pi f_\mathrm{mod} t)|$.

The observed phenomena are specific to the \textit{Resolve} instrument, but the
resonance at one of the cryocooler harmonics and a non-linear response can occur in any
instrument. The resultant power redistribution into the low-$f$ range may cause a
serious challenge for the performance of low temperature detectors regardless of whether
they are employed for calorimetric or bolometric use. We have described one such case
observed during a ground test of \textit{Resolve}. Some lessons for future missions can
be learned from this experiment: (1) a single tone sweep test may be insufficient to
understand micro-vibration phenomena, (2) tests of highly integrated systems are
required to characterize the effects, and (3) flexibility of drive frequencies may allow
to mitigate the interference at late stages of the project.

% A variable drive frequency cooler avoids resonant frequencies and may improve detector performance.
% We proposed and implemented a method to quickly and exhaustively find the optimum set from
% all combinations of refrigerator drive frequencies (256 x 32 = 8192 sets) to solve this problem in the \textit{Resolve}.

\section{Summary}\label{s5}
Here, we have given a brief description of the microphonic noise in the detector signal
caused by the cryocoolers during a ground test of the \textit{Resolve} instrument. Not
only the fundamental and harmonics of the cryocoolers driven at 15.0 and 52.0~Hz were
observed, but we also observed the low frequency ($<$20~Hz) line noise corresponding to
the beat frequency of some combinations of the harmonics of the two driving
frequencies. We have discussed that some of these effects correspond to the known
resonance frequencies of the instrument and that a non-linear behavior must be occurring
to account for power redistributed into the beat frequency tone.

\begin{acknowledgements}
 This study was made possible by the collaborative efforts of all members of the
 \textit{Resolve} team, including Sumitomo Heavy Industries, which we greatly appreciate.
\end{acknowledgements}

%\pagebreak

\end{document}